\documentclass[11pt]{article}
\usepackage[top=50mm, bottom=50mm, left=50mm, right=50mm]{geometry}
\usepackage[utf8]{inputenc}
\usepackage{siunitx}
\usepackage{lineno}
\usepackage{amssymb}
\usepackage{amsmath}
\usepackage{amsthm}
\usepackage{epsfig}
\usepackage{graphicx}
\usepackage{graphics}
\usepackage{float}
\usepackage{booktabs}
\usepackage{threeparttable}
\usepackage{subcaption}
\usepackage{multirow}
\usepackage{color}
\usepackage{lineno}
\usepackage{fullpage}
\usepackage[normalem]{ulem} 
\usepackage{makeidx}
\usepackage{xspace}
\usepackage{wrapfig}
\usepackage[capitalise]{cleveref}
\usepackage{subcaption}
\usepackage[version=4]{mhchem}
\usepackage[font=small]{caption}
\usepackage{setspace}

\providecommand{\keywords}[1]
{
  \small	
  \textbf{\textit{Keywords---}} #1
}

\usepackage{authblk}

\title{Pressure-Independent Through-Plane Electrical Conductivity Measurements of Highly Filled Conductive Polymer Composites}
\author[1,2]{Thomas Larsen}
\author[2]{Tom Larsen}
\author[2]{Søren J. Andreasen}
\author[1,*]{Jesper D. C. Christiansen}
\affil[1]{Aalborg University, Department of Materials and Production, Fibigerstræde 16, 9000 Aalborg, Denmark}
\affil[2]{Advent Technologies A/S, Lyngvej 8, 9000 Aalborg, Denmark}
\affil[*]{Corresponding author: Jesper de Claville Christiansen, jc@mp.aau.dk}

\date{}

\begin{document}

\maketitle

\doublespacing

\begin{abstract}


Highly filled conductive polymer composites (CPCs) are widely used in applications such as bipolar plate materials for polymer electrolyte membrane fuel cells and redox flow batteries, electromagnetic interference shielding and sensors due to their useful electrical properties. A common method for determining through-plane electrical conductivities $\left(\sigma_{\rm tp} \right)$ of such highly filled CPCs applies a conductive carbon paper between electrodes and sample with application of external pressure to improve electrical contact. We show the pressure-dependence of the measured $\sigma_{\rm tp}$ can be eliminated by using a liquid metal such as the gallium-indium eutectic alloy (EGaIn) as contact material. Results indicate that EGaIn reduces contact resistances and cause three to five times larger $\sigma_{\rm tp}$ compared to measurements with carbon paper contacts and pressures up to 20 bar.

\end{abstract}






\newpage

\section{Introduction}

Highly filled polymer composites with oriented and anisotropic conductive fillers are known to display an in-plane/through-plane electrical conductivity anisotropy depending on the type of filler and their orientation in the composite \cite{article:Jeong2021,article:Naji2019}. For application purposes, this anisotropy is important to quantify as current may flow along both directions. In samples of relatively high electrical conductivities, in-plane electrical conductivities $\left(\sigma_{\rm ip} \right)$ are commonly evaluated using a four-point probe technique to exclude contact resistances;
on the microscale, surfaces are rough and constitute asperities which form discrete points of contact between contacting bodies, thus constricting electrical current and resulting in a contact resistance \cite{article:Timsit1999}. The four-point probe method may also be applied to evaluate through-plane electrical conductivities $\left(\sigma_{\rm tp} \right)$, yet relatively thick samples are required for the spacing of probes \cite{article:Wei2016}. Alternatively, a two-point probe technique may be used, although the method has a significant contribution from contact resistances and thus works best for low electrical conductivity samples. Yet, a transmission line method (TLM), based on extrapolating resistances obtained at various probe distances to zero \cite{article:Kokabu2016,article:Rojo2015}, may be used in combination with the two-point probe technique to estimate the contact resistance. This approach, however, requires samples of several thicknesses which may be impractical. 

Measurements of both $\sigma_{\rm ip}$ and $\sigma_{\rm tp}$ have been widely used in the development and optimization of highly filled conductive polymer composites (CPCs) for use in redox flow batteries and polymer electrolyte membrane fuel cell applications \cite{article:Naji2019,article:Shaigan2021,article:Heinzel2004}. Such CPCs typically include high contents of conductive carbon fillers such as graphite to achieve high electrical conductivities \cite{article:Duan2021,article:Antunes2011,article:Saadat2021}. A common method for evaluating $\sigma_{\rm tp}$ of highly filled CPCs is to apply a carbon paper between the electrodes and sample under a compressive load for improved electrical contact \cite{article:Naji2019,article:Cunningham2005,article:Avasarala2009,article:Dhakate2010,article:Yao2017}. However, due to the roughness of both the composite and carbon paper, the measured $\sigma_{\rm tp}$ is dependent on the magnitude of the compressive load as well as type of carbon paper used which complicates comparison of results obtained by different laboratories \cite{article:Shaigan2021}.

Alternatively, Sow et al. \cite{article:Sow2015} used a highly filled CPC sandwiched between two current collectors with holes for sensing electrodes to directly contact the sample surface. This eliminated contact resistance between the sample and sensing electrodes. Yet, the equipotential planes close to the edge of the current collectors near the sensing electrodes bend, and thus potentials approaching the true surface potential of the sample are measured closest to the current collectors. This, however, requires fabrication of thin, concave electrodes. Q. Wei et al. \cite{article:Wei2016} employed a coaxial structure to surround their thin film samples. The outer fixed-diameter probe was the source and the diameter of the inner sense probe was varied to allow the effect of source-sense probe distance to be screened. Simulation results showed the thinner the sample, the larger the voltage drop towards the sense probe with distance away from the source. Additionally, their experimental results showed higher electrical conductivities for samples measured with the largest diameter sensing probes since potentials closer to the source probe were measured. The strongest dependence of conductivity with sensing probe diameter was observed in relatively thin samples as expected from the simulation results. This may be problematic for highly filled CPCs in battery and fuel cell applications since they typically have thicknesses on the order of millimeters.  

Instead, liquid metal contacts may be used as they are more tolerant to surface topography variations and may lower contact resistances \cite{article:Shaigan2021}. The gallium-indium eutectic alloy (EGaIn), unlike mercury, is a non-toxic liquid metal, and due to its relatively high electrical conductivity ($3.4 \times 10^4 \ \mathrm{S \ cm^{-1}}$) and shear yielding rheology it may be used to form electrical contacts \cite{article:Scharmann2004,article:Kramer2013,article:Douvogianni2018}. Due to the formation of a few nanometers thick oxide skin, EGaIn may be molded to form stable cones or droplets on surfaces \cite{article:Douvogianni2018,article:Dickey2008,article:Kim2013}.

This paper proposes a new method, independent of applied pressure, for measuring $\sigma_{\rm tp}$ by applying EGaIn as moldable electrode contacts. The accompanying setup is low-cost in terms of equipment, and it is uncomplicated to assemble. The resulting $\sigma_{\rm tp}$ obtained with this setup will be compared to the aforementioned carbon paper contact-method with partial compensation of contact resistances. 

\section{Materials and methods}

\subsection{Materials}

A commercially available woven carbon paper (stamped into discs of diameter 39 mm) and blank, grinded polymer composite plates of thickness $\sim 1.42 \ \mathrm{mm}$, consisting of a polyphenylene sulfide (PPS) matrix with graphite and carbon black filler particles, were obtained from Advent Technologies GmbH, Germany. These 'as received' plates were CNC-machined either into discs of 39 mm diameter or strips of dimensions $15 \ \mathrm{mm} \times 60 \ \mathrm{mm}$ for through-plane and in-plane electrical conductivity measurements, respectively. Gallium-indium eutectic alloy (75.5 wt.\% Ga 24.5 wt.\% In by weight, melting point 15.7 $^{\circ}$C) was procured from Merck, Germany.

To study highly filled CPCs with various surface roughness, the above-mentioned 'as received' disc-shaped samples were wet-ground with SiC paper of grit sizes P120, P320 and P1200 (European FEPA P-grit sizes). Afterwards, the samples were rinsed with de-ionized water and ethanol and dried to ensure the surface was non-contaminated. 

\subsection{Microscopy}

Scanning electron microsopy (SEM) images were acquired using a Zeiss EVO LS15 with an EHT of 15 kV. Surface topography was studied with a Zeiss Axio Imager.M2m light optical microscope (LOM). Reflected brightfield microscopy with a $50\times$ objective was used to obtain a minimum of 150 images per sample with a constant height between images. With the Zeiss ConfoMap ST software, the resulting three-dimensional stack was used to create a topographic map from which the arithmetic mean areal surface roughness $\left(S_{\rm a}\right)$ was calculated.

\subsection{Electrical conductivity}

In-plane electrical conductivity measurements were performed with a Tenma 72-2930 DC power supply and a Keysight 34465A digital multimeter. In all two- and four-point probe measurements, a current of 1 A was forced through the samples. In the two-point probe measurement with EGaIn contacts, EGaIn was initially precoated on the sample (area $A_{\rm EGaIn} = 0.13 \pm 0.01 \ \mathrm{cm^2}$) before dispensing a droplet on the coated area. The distance between EGaIn contacts was $L = 4.1 \pm 0.1 \ \mathrm{cm}$. With the four-point probe measurement, a distance between voltage-sensing probes of 2.0 cm was used. All reported in-plane electrical conductivities are based on six samples.

\begin{figure}[h]
\centering
  \includegraphics[width=.6\textwidth]{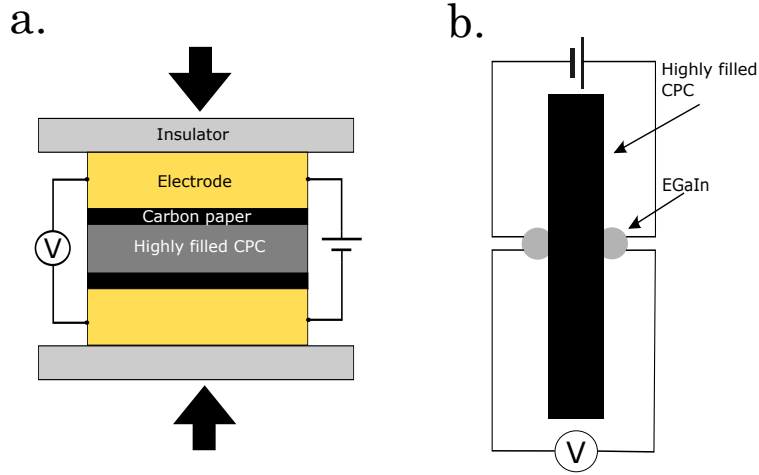}
\caption{Schematic of setups for through-plane electrical conductivity measurements on highly filled conductive polymer composites (CPCs) with carbon paper (CP) (a) and gallium-indium eutectic (EGaIn) (b) as contact material. Arrows in (a) indicate the compressive load direction.}
\label{fig:methods:setup}
\end{figure}

Through-plane electrical conductivity measurements were conducted with the same multimeter and power supply as the in-plane electrical conductivity measurements. For measurements with external pressure applied, the setup in Fig. \ref{fig:methods:setup}a with gold-coated electrodes (Au) was used. Measurements were performed at three different pressures of 4.2 bar, 12.6 bar, and 20.1 bar and three currents (1 A, 5 A, and 10 A). Lastly, to compensate partly for contact resistances between carbon paper (CP) and Au-electrodes the method described in \cite{article:Avasarala2009} was followed. The resistance measured from a single CP between the two Au-electrodes 

\begin{equation}
    R' = 2R_{\rm Au} + R_{\rm CP} + 2R_{\rm Au/CP},
    \label{eq:methods:compensate:single_GDL}
\end{equation}

with subscript '/' denoting an interface, was subtracted from the total resistance $R$ of the setup in Fig. \ref{fig:methods:setup}a

\begin{equation}
    R = 2R_{\rm Au} + 2R_{\rm CP} + 2R_{\rm Au/CP} + R_{\rm sample} + 2R_{\rm sample/CP},
    \label{eq:methods:compensate:GDL-BPP-GDL}
\end{equation}

with 'sample' denoting the highly filled CPC. In these expressions, the internal resistances in the system were assumed negligible. This leaves the compensated value

\begin{equation}
    R - R' = R_{\rm sample} + R_{\rm CP} + 2R_{\rm sample/CP}.
    \label{eq:methods:compensate:R_compensated}
\end{equation}

The through-plane electrical conductivity measurements without external pressure were conducted with the setup schematically illustrated in Fig. \ref{fig:methods:setup}b. A cotton swab was used to pre-coat EGaIn on a circular area of $0.15 \pm 0.02$ cm$^2$ on both surfaces of the sample. This was done to improve adhesion of the subsequently dispensed EGaIn droplet with the sample surface and to define the circular area through which current enters and exits the sample. Lastly, the four probes were pushed into the dispensed EGaIn droplets. The voltage was recorded at three different currents (1 A, 2.5 A, and 5 A), and the presented results are based on four to five samples. For samples tested with both the CP and EGaIn contacts, the results are based on currents of 1 A as larger currents yielded identical conductivities. 

Electrical conductivities $\left(\sigma\right)$ were calculated based on the expression $\sigma = \frac{I\cdot L}{U \cdot \mathcal{A}}$, where $I$ is the current, $U$ potential difference, $L$ separation between contacts parallel to the direction of current flow, and $\mathcal{A}$ the cross sectional area through which the current flows. In the case of EGaIn contacts, $\mathcal{A}$ was taken as the mean area of the two circular pre-coated EGaIn regions. 
\section{Results and Discussion}

\subsection{Surface roughness}

\begin{figure}[htbp]
\centering
\includegraphics[width=\textwidth]{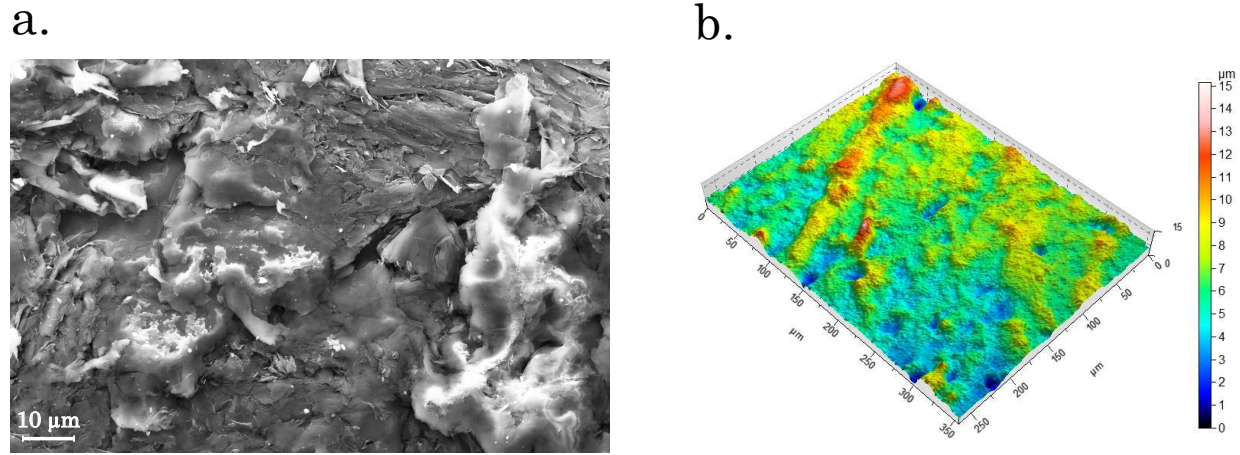}
\caption{(a) Scanning electron microscopy image of an 'as-received' highly filled conductive polymer composite surface. The white spots are likely due to charging of the surface. (b) Surface topography map of an 'as-received' highly filled conductive polymer composite sample.}
\label{fig:results:SH_grinded_SEM}
\end{figure}

The roughness of the sample surfaces was evaluated both by SEM and LOM. The resulting images of the 'as received' highly filled CPC surface are presented in \cref{fig:results:SH_grinded_SEM}, respectively.

\begin{table}[htbp]
\captionsetup{width=.8\textwidth}
\centering
\caption{Arithmetic mean areal surface roughness $\left(S_{\rm a}\right)$ for the samples used to evaluate through-plane electrical conductivities.}
\begin{threeparttable}
\label{tab:results:surface_roughness}
\begin{tabular}{lc} 
\textbf{Grit size} & \textbf{$\mathbf{S_{\rm a}}$ {[}$\mathbf{\mu}$m{]}} \\ \midrule
P120 & $6.5 \pm 2.0$ \\ \hline
P320 & $3.0 \pm 0.3$ \\ \hline
P1200 & $0.65 \pm 0.07$ \\ \hline
-\tnote{a} & $1.3 \pm 0.1$ \\ \bottomrule
\end{tabular}
   \begin{tablenotes}\footnotesize
    \item[a] 'As received' samples.
    \end{tablenotes}
\end{threeparttable}
\end{table}

Although no quantitative height information can be extracted from the SEM image, \cref{fig:results:SH_grinded_SEM}a shows a rough, flaky surface. From the topographic map, \cref{fig:results:SH_grinded_SEM}b, a height variation on the scale of several micrometers is observed, resulting in $S_{\rm a} = 1.3 \pm 0.1 \ \mu$m. The roughness and observed topography will result in a lower true contact area between CP and highly filled CPC and, thus, increase the contact resistance. 

The wet-grinding procedure produced the mean areal surface roughness presented in \cref{tab:results:surface_roughness}. As can be observed, the roughness of the grinded 'as received' samples is in between that of the samples wet-grinded with P320 and P1200 SiC paper. 

\subsection{Liquid metal contacts}

\begin{figure}[htbp]
    \centering
    \includegraphics[width=.8\textwidth]{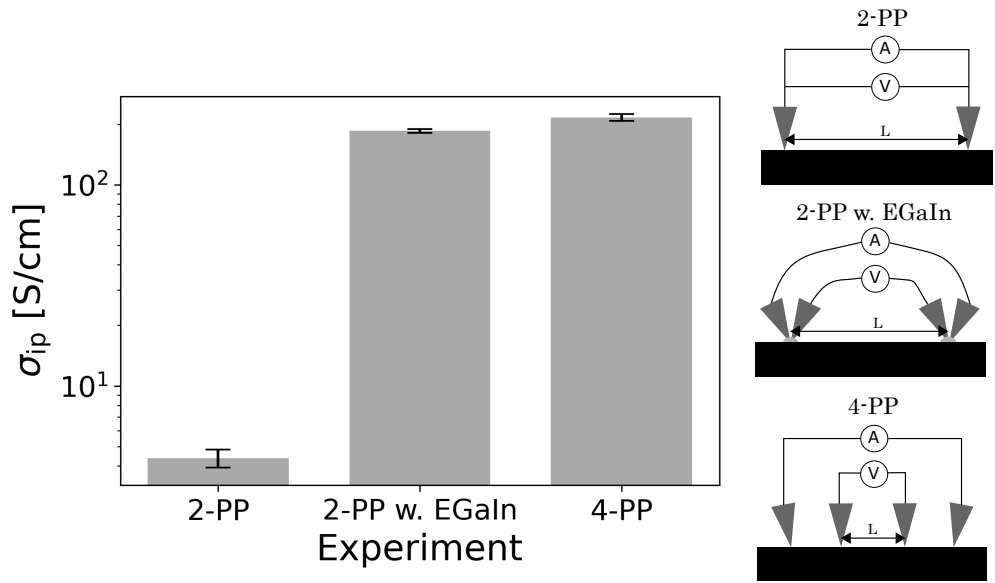}
    \caption{In-plane electrical conductivities $\left(\sigma_{\rm ip} \right)$ of 'as received' highly filled conductive polymer composites (CPCs) measured with the illustrated two- and four-point probe techniques at a current of 1 A. Error bars indicate $\pm$ one standard deviation. 2-PP: two-point probe. 2-PP w. EGaIn: two-point probe with EGaIn contacts. 4-PP: four-point probe. $L$: distance between sensing probes.}
    \label{fig:figure3}
\end{figure}

To validate the use of EGaIn as contact material for the polymer composites, measurements were conducted with both the two- and four-point probe techniques to measure in-plane electrical conductivities $\left(\sigma_{\rm ip} \right)$. A disadvantage of the two-point probe technique (2-PP in \cref{fig:figure3}) is the contact resistance between sample and electrical contacts which becomes increasingly important as the electrical conductivity of the sample increases. In the two-point probe technique with EGaIn contacts (2-PP w. EGaIn in \cref{fig:figure3}) the sample is precoated with two EGaIn contacts and both pairs of source and sensing electrodes are introduced in their respective EGaIn contacts. Hence, the measured voltage drop will include a contribution from the contact resistance between EGaIn and sample. This contact resistance is, however, eliminated with the four-point probe technique \cite{article:Rojo2015} (4-PP in \cref{fig:figure3}) used to obtain the inherent in-plane electrical conductivity of the sample. 

By comparing $\sigma_{\rm ip}$ from the 2-PP and 2-PP w. EGaIn techniques in \cref{fig:figure3}, a 42 times improvement in $\sigma_{\rm ip}$ is noticed. It is likewise interesting that $\sigma_{\rm ip}$ from 2-PP w. EGaIn and 4-PP techniques deviate by only 17 \% with the lower $\sigma_{\rm ip}$ measured by the former technique being primarily due to contact resistances between EGaIn and sample.

Writing out the contributions to the total resistance measured with the 2-PP w. EGaIn technique results in $R_{\rm 2-PP \ \mathrm{w. EGaIn}} = R_{\rm sample} + 2R_{\rm EGaIn/sample}$ with the last term accounting for the voltage drop associated with the contact resistance between EGaIn and sample. Here it has been assumed that the bulk resistance of EGaIn and any contact resistance between sensing electrode and EGaIn may be neglected due to the aforementioned high electrical conductivity of EGaIn $\left(> 10^4 \ \mathrm{S \ cm^{-1}} \right)$ and good wetting of the metal electrode by EGaIn. Finally, in the 4-PP technique, where voltage-measuring electrodes are situated on the sample of interest, it may be assumed that the measured resistance equals the sample resistance, i.e., $R_{\rm 4-PP} = R_{\rm sample}$.

The above considerations may be used to obtain an estimate of the areal contact resistance $\left(\mathcal{R}_{\rm EGaIn/sample}\right)$ between EGaIn and sample by subtracting the sample resistance from the resistance measured by the 2-PP w. EGaIn technique, i.e., $\mathcal{R}_{\rm EGaIn/sample} = R_{\rm EGaIn/sample}\cdot \left(2 A_{\rm EGaIn} \right)^{-1} = \left(R_{\rm \rm 2-PP \ \mathrm{w. EGaIn}} - R_{\rm 4-PP} \right)\cdot \left(2 A_{\rm EGaIn} \right)^{-1}$. Applying this to the investigated samples result in $\mathcal{R}_{\rm EGaIn/sample} = 2.6 \pm 0.5$ m$\Omega \ \mathrm{cm^{-2}}$. As a comparison, this value is between nine and twelve times lower than the contact resistivity of an unpolished BMC bipolar plate surface and a Toray carbon paper at pressures ranging from 0.5 - 6.5 MPa \cite{article:Avasarala2009}. This estimated areal contact resistance between EGaIn and sample will be applied to the through-plane electrical conductivity results.


\subsection{Through-plane electrical conductivity}

\begin{figure}[htbp]
    \centering
    \includegraphics[width=\textwidth]{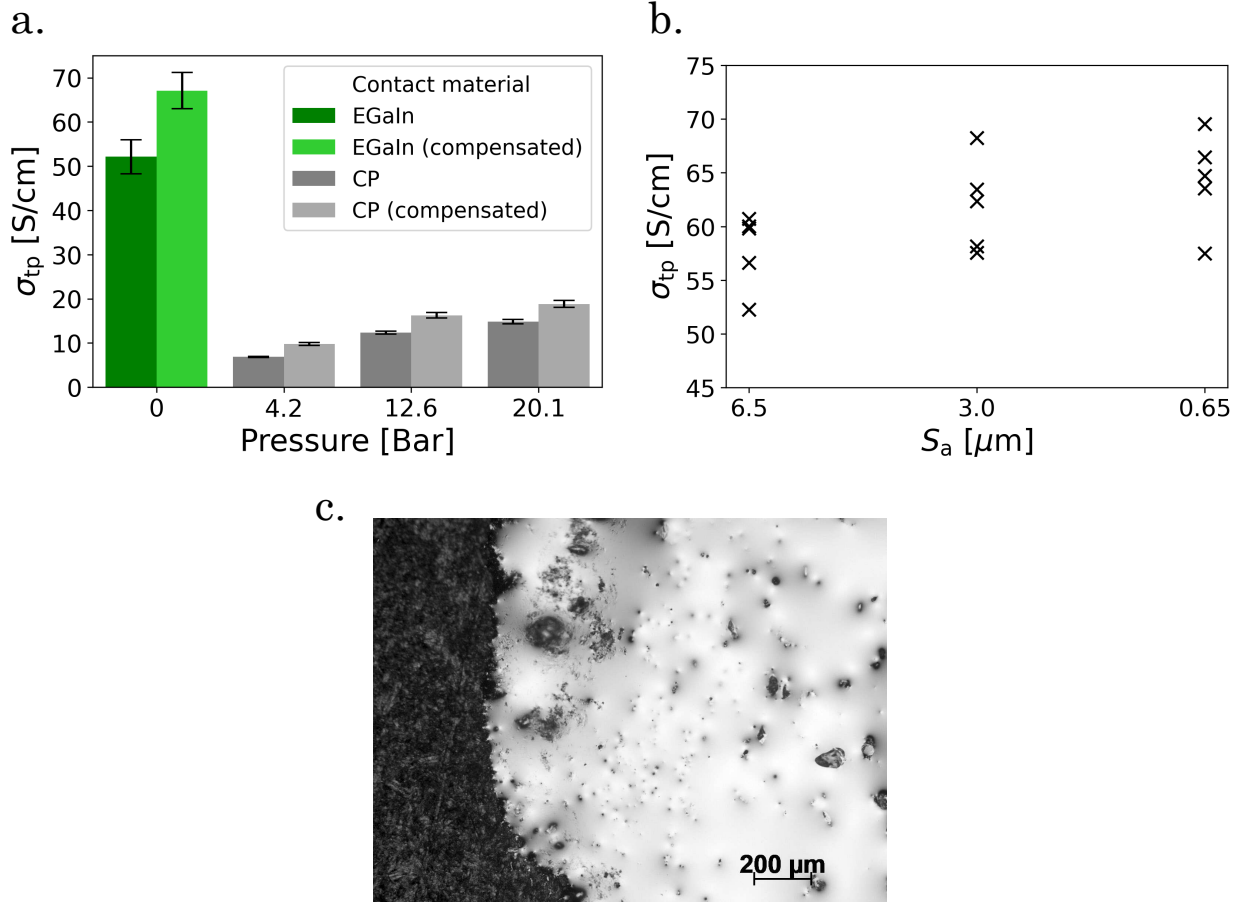}
    \caption{(a) Measured through-plane electrical conductivities $\left(\sigma_{\rm tp} \right)$ of 'as received' highly filled CPCs at a current of 1 A. Reported results are based on four and three samples for the EGaIn-method and carbon paper-method (CP), respectively. Compensated values with CP contacts are based on \cref{eq:methods:compensate:R_compensated} while with EGaIn contacts, compensated values are found by subtracting the areal contact resistance between EGaIn and highly filled CPC sample from the measured resistance. Error bars indicate $\pm$ one standard deviation. (b) Measured $\sigma_{\rm tp}$ at a current of 1 A for highly filled CPCs of various arithmetic mean areal surface roughness $\left(S_{\rm a}\right)$. The sample surface roughness and corresponding mean conductivities are: $S_{\rm a} = 6.5 \pm 2.0 \ \mu$m and $\sigma_{\rm tp} = 58 \pm 3 \ \mathrm{S \ cm^{-1}}$; $S_{\rm a} = 3.0 \pm 0.3 \ \mu$m and $\sigma_{\rm tp} = 62 \pm 4 \ \mathrm{S \ cm^{-1}}$; $S_{\rm a} = 0.65 \pm 0.07 \ \mu$m and $\sigma_{\rm tp} = 64 \pm 4 \ \mathrm{S \ cm^{-1}}$. Five different samples were tested at each surface roughness. (c) Light optical microscopy image of an 'as received' highly filled CPC (dark grey) pre-coated with EGaIn (silver). The dark spots are regions not wetted by EGaIn.}
    \label{fig:figure4}
\end{figure}

The measured through-plane electrical conductivities obtained with either CPs or EGaIn as contact material are presented in \cref{fig:figure4}a. With CPs, the compensated $\sigma_{\rm tp}$ (see \cref{eq:methods:compensate:R_compensated}) is seen to almost double from $10 \ \mathrm{S \ cm^{-1}}$ at 4.2 bar to $19 \ \mathrm{S \ cm^{-1}}$ at 20.1 bar. The conductivity increase with pressure results mainly from the decrease in interfacial resistance between the CP and sample due to an increased effective contact area caused by deformation of surface asperities on the CP and sample surfaces \cite{article:Sow2015}. Observing the discrepancy between the compensated and uncompensated $\sigma_{\rm tp}$, a decreased relative deviation from 42 \% at 4.2 bar to 27 \% at 20.1 bar is noted. To some degree, this is influenced by an increased effective contact area between CP and electrodes, but the resistivity of CPs also depend on compressive pressure due to the formation of fiber cracks and a resulting alteration of conductive pathways \cite{article:Qiu2018}.

\cref{fig:figure4}a indicates an up to six times higher measured $\sigma_{\rm tp}$ using EGaIn as contact material in contrast to CP contacts. Besides the comparatively higher conductivity of EGaIn compared to CP, this could be due to EGaIn more easily conforming to the surface topography of the sample than the solid CP material, thereby increasing the effective contact area and hence reducing the contact resistance \cite{article:Shaigan2021}.

Since the contact resistance between EGaIn and highly filled CPC sample is normal to the sample surface, and the difference between the approach in \cref{fig:methods:setup}b and the 2-PP w. EGaIn technique lies solely in the current flow direction through the sample, the areal contact resistance $\mathcal{R}_{\rm EGaIn/sample}$ from in-plane electrical conductivity measurements can be used to compensate for contact resistance in $\sigma_{\rm tp}$-measurements. Hence, $\mathcal{R}_{\rm EGaIn/sample}$ is multiplied by the area of EGaIn contacts from $\sigma_{\rm tp}$-measurements to obtain a representative contact resistance contribution to $\sigma_{\rm tp}$. Subtracting this resistance from the total resistance measured with the setup in \cref{fig:methods:setup}b for 'as received' samples results in $\sigma_{\rm tp}^{\rm comp} = 67 \pm 4 \ \mathrm{S \ cm^{-1}}$, a value 29 \% larger than the uncompensated $\sigma_{\rm tp}^{\rm uncomp} = 52 \pm 4 \ \mathrm{S \ cm^{-1}}$, \cref{fig:figure4}a.  

\cref{fig:figure4}b shows the variation of $\sigma_{\rm tp}$ with surface roughness of the highly filled CPC surface. Overlap of $\sigma_{\rm tp}$ at all three roughness-values is observed. The lower values of $\sigma_{\rm tp}$ obtained for surfaces with $S_{\rm a} = 6.5 \pm 2.0 \ \mu$m suggest a greater contact resistance due to poorer wetting of the rough surface. As can be seen in \cref{fig:figure4}c, precoating with EGaIn does not ensure uniform wetting of the substrate. It can be expected that the lower surface energy of PPS compared to the conductive fillers causes EGaIn to preferably wet the latter. Thus, returning to the SEM image, \cref{fig:results:SH_grinded_SEM}, the white spots could indicate local areas of PPS at the surface which may hinder uniform wetting of the substrate. Additionally, with a head diameter on the order of millimeters, the cotton swabs used for precoating the substrate are expected to have more difficulty distributing the EGaIn in valleys of relatively rough surfaces, thereby offering a possible explanation for the trend observed in \cref{fig:figure4}b. So, the proposed method appears less suited to study rough surfaces. Yet, highly filled CPCs for fuel cell bipolar plate applications in general benefit from possessing smooth surfaces due to lower contact resistance and, hence, lower voltage drop \cite{article:San2017}. Thus, the proposed approach may be a viable alternative to characterize $\sigma_{\rm tp}$ for said applications.  

\section{Conclusion}

Measurements of through-plane electrical conductivities $\left(\sigma_{\rm tp}\right)$ on highly filled conductive polymer composites using either (i) carbon papers as contact material with externally applied pressure, or (ii) a liquid gallium-indium eutectic alloy (EGaIn), result in higher measured $\sigma_{\rm tp}$ by adopting the latter method. Thus, with EGaIn as electrical contacts a decrease in contact resistance is observed, resulting in $\sigma_{\rm tp}$-values more closely resembling those of the composites. Furthermore, as the method of EGaIn as contact material is pressure-independent, comparison of experimental results between laboratories is expected to become more straightforward. 

It would be fruitful to investigate the applicability of alternative metals to EGaIn that, similarly, are liquid at room temperature, e.g., GaInSn eutectic alloy ('Galinstan') \cite{article:Scharmann2004}. An alternative approach could be to use conductive colloidal gels with conductive granular fillers such that the contacts possess a yield stress sufficiently high to allow forming stable contacts and prevent sedimentation of the granular fillers.
\section*{Conflicts of interest}

The authors declare they have no conflicts of interest. 

\section*{Acknowledgements}

This work was supported in part by Advent Technologies A/S and a grant from the Industrial PhD programme, Innovation Fund Denmark, project 8053-00063B.

\newpage

\bibliographystyle{ieeetr}

\begin{thebibliography}{10}

\bibitem{article:Jeong2021}
K.~I. Jeong, J.~Oh, S.~A. Song, D.~Lee, D.~G. Lee, and S.~S. Kim, ``{A review
  of composite bipolar plates in proton exchange membrane fuel cells:
  Electrical properties and gas permeability},'' {\em Compos. Struct.},
  vol.~262, p.~113617, 2021.

\bibitem{article:Naji2019}
A.~Naji, B.~Krause, P.~Potschke, and A.~Ameli, ``{Hybrid conductive
  filler/polycarbonate composites with enhanced electrical and thermal
  conductivities for bipolar plate applications},'' {\em Polym. Compos.},
  vol.~40, pp.~3189--3198, 2019.

\bibitem{article:Timsit1999}
R.~Timsit, ``Electrical contact resistance: properties of stationary
  interfaces,'' {\em IEEE Trans. Compon. Packag. Technol.}, vol.~22, no.~1,
  pp.~85--98, 1999.

\bibitem{article:Wei2016}
Q.~Wei, H.~Suga, I.~Ikeda, M.~Mukaida, K.~Kirihara, Y.~Naitoh, and T.~Ishida,
  ``{An accurate method to determine the through-plane electrical conductivity
  and to study transport properties in film samples},'' {\em Org. Electron.},
  vol.~38, pp.~264--270, 2016.

\bibitem{article:Kokabu2016}
T.~Kokabu, S.~Inoue, and Y.~Matsumura, ``{Estimation of adsorption energy for
  water molecules on a multi-walled carbon nanotube thin film by measuring
  electric resistance},'' {\em AIP Advances}, vol.~6, p.~115212, 2016.

\bibitem{article:Rojo2015}
M.~M. Rojo, C.~V. Manzano, D.~Granados, M.~R. Osorio, T.~Borca-Tasciuc, and
  M.~Martin-Gonzalez, ``{High electrical conductivity in out of plane direction
  of electrodeposited Bi$_2$Te$_3$ films},'' {\em AIP Advances}, vol.~5,
  p.~087142, 2015.

\bibitem{article:Shaigan2021}
N.~Shaigan, X.-Z. Yuan, F.~Girard, K.~Fatih, and M.~Robertson, ``{Standardized
  testing framework for quality control of fuel cell bipolar plates},'' {\em J.
  Power Sources}, vol.~482, p.~228972, 2021.

\bibitem{article:Heinzel2004}
A.~Heinzel, F.~Mahlendorf, O.~Niemzig, and C.~Kreuz, ``{Injection moulded low
  cost bipolar plates for PEM fuel cells},'' {\em J. Power Sources}, vol.~131,
  pp.~35--40, 2004.

\bibitem{article:Duan2021}
Z.~Duan, Z.~Qu, Q.~Ren, and J.~Zhang, ``{Review of bipolar plate in redox flow
  batteries: Materials, structures, and manufacturing},'' {\em Electrochem.
  Energy Rev.}, vol.~4, pp.~718--756, 2021.

\bibitem{article:Antunes2011}
R.~A. Antunes, M.~C.~L. de~Oliveira, G.~Ett, and V.~Ett, ``{Carbon materials in
  composite bipolar plates for polymer electrolyte membrane fuel cells: A
  review of the main challenges to improve electrical performance},'' {\em J.
  Power Sources}, vol.~196, pp.~2945--2961, 2011.

\bibitem{article:Saadat2021}
N.~Saadat, H.~N. Dhakal, J.~Tjong, S.~Jaffer, W.~Yang, and M.~Sain, ``{Recent
  advances and future perspective of carbon materials for fuel cell},'' {\em
  Renew. Sust. Energ. Rev.}, vol.~138, p.~110535, 2021.

\bibitem{article:Cunningham2005}
N.~Cunningham, M.~Lefevre, G.~Lebrun, and J.-P. Dodelet, ``{Measuring the
  through-plane electrical resistivity of bipolar plates (apparatus and
  methods)},'' {\em J. Power Sources}, vol.~143, pp.~93--102, 2005.

\bibitem{article:Avasarala2009}
B.~Avasarala and P.~Haldar, ``{Effect of surface roughness of composite bipolar
  plates on the contact resistance of a proton exchange membrane fuel cell},''
  {\em J. Power Sources}, vol.~188, pp.~225--229, 2009.

\bibitem{article:Dhakate2010}
S.~R. Dhakate, S.~Sharma, N.~Chauhan, R.~K. Seth, and R.~B. Mathur, ``{CNTs
  nanostructuring effect on the properties of graphite composite bipolar
  plate},'' {\em Int. J. Hydrog. Energy}, vol.~35, pp.~4195--4200, 2010.

\bibitem{article:Yao2017}
K.~Yao, D.~Adams, A.~Hao, J.~P. Zheng, Z.~Liang, and N.~Nguyen, ``{Highly
  conductive and strong graphite-phenolic resin composite for bipolar plate
  applications},'' {\em Energy Fuels}, vol.~31, pp.~14320--14331, 2017.

\bibitem{article:Sow2015}
P.~K. Sow, S.~Prass, P.~Kalisvaart, and W.~Merida, ``{Deconvolution of
  electrical contact and bulk resistance of gas diffusion layers for fuel cell
  applications},'' {\em Int. J. Hydrog. Energy}, vol.~40, pp.~2850--2861, 2015.

\bibitem{article:Scharmann2004}
F.~Scharmann, G.~Cherkashinin, V.~Breternitz, C.~Knedlik, G.~Hartung, T.~Weber,
  and J.~A. Schaefer, ``{Viscosity effect on GaInSn studied by XPS},'' {\em
  Surf. Interface Anal.}, vol.~36, pp.~981--985, 2004.

\bibitem{article:Kramer2013}
R.~K. Kramer, J.~W. Boley, H.~A. Stone, J.~C. Weaver, and R.~J. Wood, ``{Effect
  of microtextured surface topography on the wetting behavior of eutectic
  gallium-indium alloys},'' {\em Langmuir}, vol.~30, pp.~533--539, 2013.

\bibitem{article:Douvogianni2018}
E.~Douvogianni, X.~Qiu, L.~Qiu, F.~Jahani, F.~B. Kooistra, J.~C. Hummelen, and
  R.~C. Chiechi, ``{Soft nondamaging contacts formed from eutectic Ga-In for
  the accurate determination of dielectric constants of organic materials},''
  {\em Chem. Mater.}, vol.~30, pp.~5527--5533, 2018.

\bibitem{article:Dickey2008}
M.~D. Dickey, R.~C. Chiechi, R.~J. Larsen, E.~A. Weiss, D.~A. Weitz, and G.~M.
  Whitesides, ``{Eutectic gallium-indium (EGaIn): A liquid metal alloy for the
  formation of stable structures in microchannels at room temperature},'' {\em
  Adv. Funct. Mater.}, vol.~18, pp.~1097--1104, 2008.

\bibitem{article:Kim2013}
D.~Kim, P.~Thissen, G.~Viner, D.-W. Lee, W.~Choi, Y.~J. Chabal, and J.-B. Lee,
  ``{Recovery of nonwetting characteristics by surface modification of
  gallium-based liquid metal droplets using hydrochloric acid vapor},'' {\em
  ACS Appl. Mater. Interfaces}, vol.~5, pp.~179--185, 2013.

\bibitem{article:Qiu2018}
D.~Qiu, H.~Jan{\ss}en, L.~Peng, P.~Irmscher, X.~Lai, and W.~Lehnert,
  ``{Electrical resistance and microstructure of typical gas diffusion layers
  for proton exchange membrane fuel cell under compression},'' {\em Appl.
  Energy}, vol.~231, pp.~127--137, 2018.

\bibitem{article:San2017}
F.~G.~B. San and O.~Okur, ``{The effect of compression molding parameters on
  the electrical and physical properties of polymer composite bipolar
  plates},'' {\em Int. J. Hydrog. Energy}, vol.~42, pp.~23054--23069, 2017.

\end{thebibliography}

\end{document}